# NEW MEASUREMENT OF $\frac{\Delta G}{G}$ AT COMPASS

S.L. PROCUREUR, on behalf of the COMPASS collaboration
*DSM/DAPNIA/SPhN, CEA-Saclay, Orme des merisiers,
91191 Gif sur Yvette, France*

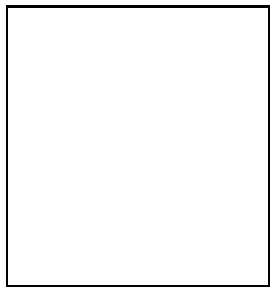

One of the main goals of the COMPASS experiment at CERN is the determination of the gluon polarisation in the nucleon, $\Delta G/G$. It is extracted from the spin asymmetry in the scattering of a polarized 160 GeV muon beam on a polarized LiD target, by selecting photon-gluon fusion events. These events are tagged by the production of open charm or high $p_T$ hadron pairs. We present the results obtained for $\frac{\Delta G}{G}(x_G)$ in both channels. For the first time, preliminary results of 2004 data in the high $p_T$ channel are also shown.

## 1 General overview

The decomposition of the nucleon spin in terms of the contributions of its constituents has been a central topic in polarized deep-inelastic scattering (DIS) experiments for the last 20 years. It can be written as:

$$\frac{1}{2} = \frac{1}{2}\Delta\Sigma + \Delta G + L_q + L_g, \qquad (1)$$

where $\frac{1}{2}\Delta\Sigma$ is the contribution from the spin of quarks, $\Delta G$ the contribution from the spin of the gluons, and $L_q + L_g$ is the orbital angular momentum of quarks and gluons.

The first exciting result was obtained by the EMC[1] experiment at CERN, which showed that the spin of the quarks only contributes to a small fraction of the proton spin. Other experiments confirmed this result, establishing $a_0 \sim \Delta\Sigma$ between 20 and 30%, i.e. in contradiction with the 60% expected in the quark-parton model. However, due to axial anomaly, $a_0 = \Delta\Sigma - \frac{3\alpha_s}{2\pi}\Delta G$ in some schemes, so $\Delta\Sigma$ can still be large if the contribution from the spin of gluons is large enough. Up to recently, $\Delta G$ could only be determined in inclusive DIS, using the $Q^2$ dependence of the polarized structure function $g_1$. Unfortunately, the precision of such a determination is strongly limited by the small $Q^2$ range of existing data. Several experiments therefore concentrated on a direct determination of the gluon polarization $\frac{\Delta G}{G}$: STAR and PHENIX at the RHIC p-p collider, and the fixed target experiments HERMES and COMPASS.

COMPASS is a polarized DIS experiment installed on the M2 beam line of the CERN SPS which delivers a 160 GeV polarized muon beam with an intensity of $2 \cdot 10^8$ muons per SPS cycle (i.e. every 16s). The polarized $^6$LiD (deuteron) target is split in an upstream and a downstream cell with opposite polarizations. The position and the momentum of the incoming muons are measured via a set of trackers and a beam momentum station located upstream of the target. The particles produced by the $\mu N$ interactions are detected downstream of the target in a double-stage spectrometer with high rate capability, containing around 200 tracker planes.

## 2   Determination of $\frac{\Delta G}{G}(x_G)$

In polarized semi-inclusive DIS, a direct measurement of the polarization $\Delta G/G(x_G)$ of the gluons carrying a fraction $x_G$ of the nucleon momentum can be obtained from the cross-section helicity asymmetry $A_{\parallel}$ of an event sample enriched with photon-gluon fusion (PGF) processes, $\gamma^* g \to q\bar{q}$. Indeed, assuming factorization:

$$A_{\parallel} = R_{PGF} \hat{a}_{LL}^{PGF} \frac{\Delta G}{G} + A_{backgd}, \qquad (2)$$

where $\hat{a}_{LL}^{PGF}$ is the helicity asymmetry of the hard lepton-gluon scattering, $R_{PGF}$ the fraction of PGF events in the selected sample, and $A_{backgd}$ the asymmetry of background processes.
Two different methods are used at COMPASS to tag the PGF: one consists in selecting charmed particles, providing a pure but small sample of PGF events; the other selects events containing 2 hadrons with high transverse momentum $p_T$ with respect to the virtual photon direction.

### 2.1   Open charm channel

Due to its large mass ($\sim$1.5 GeV/c$^2$), it is very unlikely to find a $c$ quark in the nucleon, or to create it during the fragmentation. Moreover, at tree level, $c$ quarks can only be produced in the photon-gluon fusion process. That is why the detection of a charmed hadron in the final state is a very clear signature of PGF events. Also it was checked on data that the remaining (combinatorial) background does not contribute to the asymmetry, $A_{backgd} \approx 0$. The selection is done by reconstructing $D^0$ or $D^*$ from the decays: $D^0 \to K\pi$ or $D^* \to D^0 \pi_s \to K\pi\pi_s$, combinatorial background being largely suppressed in the second case due to the small available phase space for the soft pion.
A preliminary analysis of 2002 and 2003 data gives:

$$\frac{\Delta G}{G}(x_G \sim 0.15) = -1.08 \pm 0.073 (stat) \qquad (3)$$

The analysis of 2004 data is in progress, leading to an expected statistical error of 0.43.

### 2.2   High $p_T$ hadron pair channel

In this case, the statistics is much larger, but a non-negligible fraction of background processes is still present in the selected sample, $A_{backgd} \neq 0$, and the fraction of PGF events is unknown. These quantities can only be estimated using a Monte Carlo simulation, thus introducing a model dependence in our determination of $\Delta G/G$.
For the first time, 2004 data are included in this analysis. The selected events are required to contain at least 2 charged hadrons associated to the primary vertex, in addition to the incident and scattered muons. Since the generator we use, PYTHIA, provides a reliable model for interactions of virtual photons with nucleons in the low virtuality region [2], we select events with $Q^2 < 1$ (GeV/c)$^2$. For events with $Q^2 > 1$ (GeV/c)$^2$ (10% of the statistics), a separate analysis is performed using the LEPTO generator. To enhance the fraction of PGF events in our sample, we require the transverse momentum of the 2 hadrons to be large: $p_T^{h1} > 0.7$ GeV/c, $p_T^{h2} > 0.7$ GeV/c and $(p_T^{h1})^2 + (p_T^{h2})^2 > 2.5$ (GeV/c)$^2$, as in the SMC high $p_T$ analysis[3]. In total, around 500,000 events remain after these cuts, defining our high-$p_T$ sample, from which we extract the following asymmetry (defined in [4]):

$$\left\langle \frac{A_\parallel}{D} \right\rangle = +0.004 \pm 0.013(stat) \pm 0.003(syst). \qquad (4)$$

The systematic error accounts for the false asymmetries related to the apparatus. Other sources of systematics errors, including the error on the beam and target polarizations, are proportional to the (small) measured asymmetry, and have been therefore neglected.

As already stated, our determination of the gluon polarization involves a Monte Carlo simulation. The events generated by PYTHIA are propagated through a GEANT model of our apparatus, and reconstructed with the same program as for real data. The cuts described above are then applied, and the selected events define our Monte Carlo high $p_T$ sample.

In our kinematics, PYTHIA generates 2 kinds of processes. In the *direct processes*, the virtual photon directly takes part in the hard partonic interaction, whereas in the *resolved processes*, it first fluctuates to a hadronic state, from which a parton is extracted and interacts with a parton from the nucleon. These resolved processes account for nearly 60% of our Monte Carlo high $p_T$ sample.

To reach a good agreement between our Monte Carlo and the data, we varied many parameters of the generator, and we finally found that only one should be changed, i.e the width of the intrinsic transverse momentum distribution of partons in the resolved virtual photon was decreased from 1 GeV/c to 0.5 GeV/c. Fig. 1 presents a comparison between the Monte Carlo and the real data high $p_T$ samples.

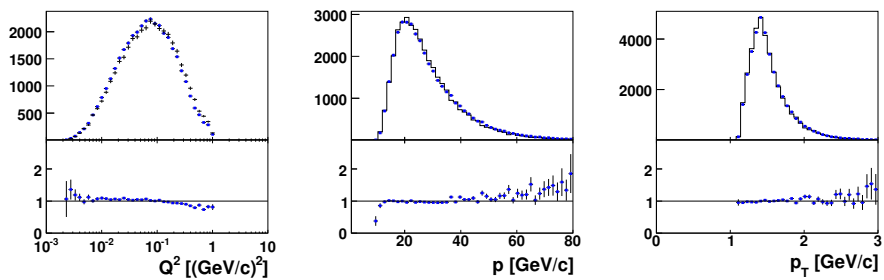

Figure 1: The upper parts of these plots show a comparison between the simulated (histograms) and real data (points) samples of high $p_T$ events, normalized to the number of events. The lower part shows the corresponding data/simulation ratio. $p$ ($p_T$) is the total (transverse) momentum of the leading hadron. A similar agreement is obtained for the next-to-leading hadron.

Taking into account all the generated subprocesses, the high $p_T$ asymmetry can be approximately expressed as:

$$\left\langle \frac{A_\parallel}{D} \right\rangle = R_{PGF} \left\langle \hat{a}_{LL}^{PGF}/D \right\rangle \frac{\Delta G}{G} + R_{QCDC} \left\langle \hat{a}_{LL}^{QCDC}/D \right\rangle A_1 \\ + \sum_{f,f'=u,d,s,\bar{u},\bar{d},\bar{s},g} R_{ff'} \left\langle \hat{a}_{LL}^{ff'} \left(\frac{\Delta f}{f}\right)^d \left(\frac{\Delta f'}{f'}\right)^\gamma \right\rangle \qquad (5)$$

The contribution of the PGF events to the high $p_T$ asymmetry was found to be $-0.292 \times \Delta G/G$ and the contribution of QCD Compton $0.0063$ [5], using a fit on the world data for the virtual photon deuteron asymmetry $A_1$. For resolved photon processes, some of them involve a gluon in the nucleon, and are thus part of the signal. Their analysing powers $\hat{a}_{LL}^{ff'}$ can be calculated in pertubative QCD [6], and the polarizations $(\Delta f/f)^d$ of the u, d and s quarks in the deuteron were estimated using the GRV98 [7] and GRSV2000 [8] parameterizations. On the other hand, the polarized PDFs of the virtual photon are presently unknown. However, theoretical considerations [9] provide a minimum and a maximum value for each $(\Delta f')^\gamma$. We were thus able to estimate the contribution of all the resolved photon processes to be in some ($\Delta G/G$ dependent) range.

Finally, the systematic errors related to the use of PYTHIA were estimated by varying the relevant parameters of this generator in a range where the agreement between our simulation and the data remains reasonable. It appeared that the result for $\Delta G/G$ depends essentially on the width of the distribution of the intrinsic transverse momentum for partons in the resolved photon: the variation of this parameter results indeed in a variation of 30% for $R_{PGF}$.

Gathering all the pieces, and using the preliminary asymmetries shown in Fig. 2(left), we obtain:

$$\frac{\Delta G}{G}(x_G = 0.085, \mu^2 \approx 3(GeV/c)^2) = 0.016 \pm 0.058(stat) \pm 0.055(syst) \qquad (6)$$

The nucleon momentum fraction $x_G$ carried by the gluon, as well as the hard scale $\mu^2$ on which $\Delta G/G$ depends weakly, were estimated using the simulation. Fig. 2(right) shows a comparison of our result with previous measurements and recent GRSV parameterizations, resulting from QCD fits to the world $g_1$ data.

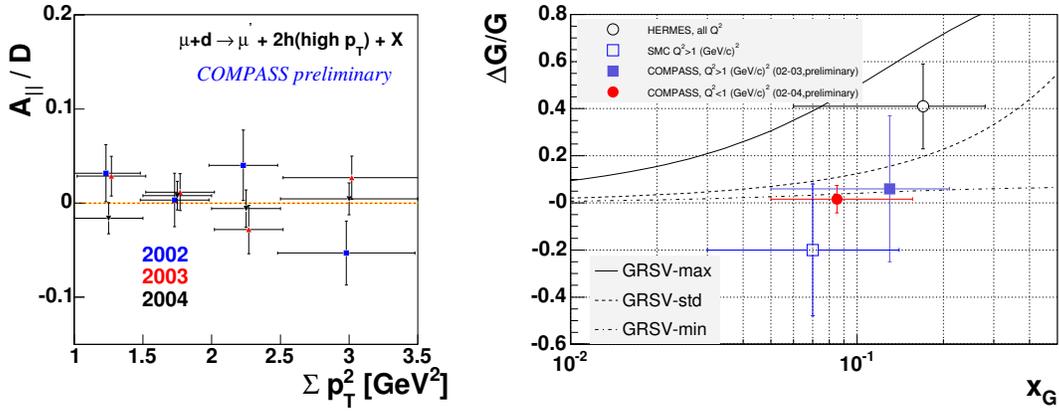

Figure 2: (left): measured asymmetry $A_{\parallel}/D$ with the high $p_T$ sample, for the 3 years of data taking, as a function of $(p_T^{h1})^2 + (p_T^{h2})^2$. Only the last bin is used for the extraction of $\Delta G/G$; (right): comparison of our result with previous measurements and NLO GRSV parameterizations at $\mu^2 = 3$ $(GeV/c)^2$, corresponding to $\Delta G = 2.5$ (max), 0.6 (std) and 0.2 (min).

Values between 0 and 0.6 for $\Delta G$ are clearly favored, unless $\Delta G(x_G)$ has a node around $x_G \sim 0.1$. The use of 2006 data, as well as a neural network for event selection and an extraction of $\frac{\Delta G}{G}(x_G)$ for different $x_G$ should allow us to better constrain $\Delta G$ in a near future.